\documentclass[prb,showpacs, aps,twocolumn]{revtex4}

\usepackage{graphicx}
\usepackage{amsmath}
\usepackage{amssymb}

\begin{document}

\title{Time-dependent magneto-transport in a driven graphene spin valve}
\author {Kai-He Ding$^{1,2}$, Zhen-Gang Zhu$^2$, and Jamal Berakdar$^2$}
\affiliation{$^1$Department of Physics and Electronic Science,
 Changsha University of Science and Technology,
 Changsha,410076, China\\
 $^2$Institut f\"{u}r Physik, Martin-Luther-Universit\"{a}t
Halle-Wittenberg,  06099 Halle (Saale), Germany}

\begin{abstract}
Based on the time-dependent nonequilibrium Green's function method
we investigate  theoretically the time and spin-dependent
transport through a graphene layer upon the application of a
static  bias voltage to the electrodes and a  time-alternating
 gate voltage to  graphene. The  electrodes  are magnetic with
  an arbitrary mutual orientations of their magnetizations.
We find features in the current that are governed by
an interplay of the strength of the  alternating field  and the
Dirac point in graphene: The influence of a
 weak alternating field  on the zero bias
conductance is strongly suppressed by the zero density of
state at the Dirac point.   In contrast, for  a
strong amplitude of the alternating field  the
 current is dominated by several resonant peaks, in particular
   a marked peak appears at zero bias.
This subtle competition  results in a transition of the
tunnel magnetoresistance from a broad peak to a sharp dip at
a zero bias voltage applied to the electrodes. The dip amplitude can be manipulated by
tuning the ac field frequency.
\end{abstract}
 \maketitle

\section{Introduction}
The discovery of  graphene, a single layer of carbon, sparked a
burst of research  resulting in a variety  of
fascinating findings \cite{novoselov2,zhang,geim,castrocond}.
 The modification of the electronic properties of graphene under various external
conditions have also been explored extensively. Particularly interesting examples  are the
response to electromagnetic fields and the
 frequency dependence of the
conductivity \cite{peres,falkepjb2007,gusyprb2006}, the photon-assisted
transport\cite{trauprb2007} and current\cite{mosk}, the microwave and far-infrared
response\cite{gusyprb2007,falkprb2007,aberprb2007}, as well as the plasmon
spectrum\cite{hwangprb2007,wunsnjp2006,apalijm2007}. In addition, new
electromagnetic modes akin to graphene  have   been
predicted\cite{vafeprl2006,mikhprl2006} and it was argued that graphene
responds intrinsically in a
non-linear manner  to electromagnetic radiations and may serve thus as
a  natural frequency multiplier with potential
 applications  in terahertz
electronics \cite{mikjpcm2008}.

 Recently, considerable attention was devoted
 to  the spin-dependent transport in  graphene and graphene-ferromagnet
 heterostructures, as prototypical spintronic systems
 \cite{hill2006ieee,tombros2007nature,cho2007apl,ohishi2007apl,wang2008prb,ding2009prb,maassen,dingall}.
Several groups have been successful in experimentally  contacting
graphene to ferromagnetic electrodes \cite{maassen} such as cobalt
electrodes\cite{tombros2007nature,ohishi2007apl} and permalloy
electrodes\cite{cho2007apl}. This rendered possible  the demonstration of
spin injection into a graphene thin film by means of  nonlocal
magnetoresistance measurements. Further experiments evidenced  a
rather long spin-flip relaxation length $\approx 1 \mu m$
 in a single layer graphene even at room
temperatures\cite{tombros2007nature}.  An anomalous
cusp-like feature of the magnetoresistance versus the applied bias
was observed in a graphene-based spin-valve devices\cite{wang2008prb}.
Following these
works, Ding \emph{et al.}\cite{ding2009prb} studied theoretically
the spin-dependent transport through the graphene spin valve
device, and pointed out that the cusp-like feature at zero
bias is due to  a subtle interplay  of the
graphene peculiar nature and the conventional spin-valve properties. More
recently, the effects of the external conditions such as
disorder\cite{chenjpcm2009} and the magnetic
impurities\cite{ding}, on the spin dependent transport in
graphene-ferromagnet
 heterostructures have been addressed.

 On the other hand, the spin-dependent transport under an alternating (ac) field is relatively less explored.
 In this work, we study theoretically the spin-dependent transport through a graphene spin-valve device
 with statically biased ferromagnetic leads having arbitrary spin-polarization
directions in the presence of an external ac field acting on the graphene monolayer
in form of a gate voltage.
The employed method is based on the time
dependent nonequilibrium Green's function approach,
 as described in  Refs.\onlinecite{haug,rammer}.
It is found that the current shows a peculiar behaviour when the strength of the
alternating field is varied.  For a small amplitude of the ac field,
the zero DOS at  the Dirac point of graphene suppresses
strongly the ac field effect, particularly the  peak at zero bias
is diminished.  When the ac field
strength becomes sufficiently large, the insulator-like properties
of graphene  at the Dirac point are less relevant and  a prominent peak
appears  in the differential conductance at
 zero bias. In this case, the TMR exhibits a transition
from a peak to a sharp dip at a zero bias voltage. The dip magnitude can  be
varied  by changing the ac field frequency.

%
%
%
%
%
%
%
\section{Theoretical model}
The spin valve device under consideration here
 consists of a single layer of graphene
sandwiched between two ferromagnetic
electrodes. The magnetization  $\mathbf{M}_L$ of the left electrode is
assumed to align along the $y$ direction, while that of the right electrode
$\mathbf{M}_R$   deviates from the $y$
direction by a relative angle $\theta$.
A dc  bias  $V$ applied to the electrodes results
in a longitudinal static current in graphene.
In addition, we consider a harmonic ac gate  voltage $V_{g}(t)$  with frequency $\omega$
and strength $V_{ac}$ 
being applied to the graphene sheet, i.e.
\begin{equation} V_{g}(t) =V_{ac}\cos\omega t .\label{eq:ac}\end{equation}
The nature of the coupling of $V_{g}(t)$ to the graphene sheet follows from the symmetry group analysis,
e.g. as done recently in Ref. [\onlinecite{winkler}] in full details.
Inspecting the results of Ref. \cite{winkler} we conclude that $V_{g}(t)$
couples to the graphene sheet through the $\Gamma_{1}$ representation of the point group of $D_{3h}$
at the Dirac points of graphene. In effect, $V_{g}(t)$ acts on the transport as ac Fermi level.
Nevertheless, as shown below,  this simple coupling leads to remarkable consequences for the transport
and TMR\cite{footn}.
 In the present case the Hamiltonian of this system reads thus
\begin{equation}
H=H_L+H_R+H_G+H_T
\end{equation}
where
\begin{equation}
H_L=\sum\limits_{\mathbf{k},\sigma}
\varepsilon_{\mathbf{k}L\sigma} c_{\mathbf{k}L\sigma}^\dag
c_{\mathbf{k}L\sigma}
\end{equation}
\begin{equation}
H_{R}=\sum\limits_{\mathbf{k},\sigma }(\varepsilon _{\mathbf{k}R}-\sigma \mathbf{M}%
_{R}\cos \theta )c_{\mathbf{k}R\sigma }^{\dag
}c_{\mathbf{k}R\sigma }-\mathbf{M}_{R}\sin \theta
c_{\mathbf{k}R\sigma }^{\dag }c_{\mathbf{k}R\overline{\sigma }}
\label{03}
\end{equation}
where $ \varepsilon_{\textbf{k}\alpha\sigma}$ is the single
electron energy associated with the momentum $\mathbf{k}$ and the
spin $\sigma$ in the $\alpha=L,R$ electrode,
$c_{\textbf{k}\alpha\sigma}^\dag(c_{\textbf{k}\alpha\sigma})$ is
the usual creation (annihilation) operator for an electron with
the energy $\varepsilon_{\mathbf{k}\alpha\sigma}$.
%

Having specified the nature of the coupling of $V_{g}$ to graphene we write for the
graphene Hamiltonian in
the tight-binding approximation
\begin{equation}
H_G=\sum\limits_{i,\sigma}\epsilon(t)(a_{i,\sigma}^\dag
a_{i,\sigma}+b_{i,\sigma}^\dag b_{i,\sigma})-t_{\text{g}}\sum\limits_{\langle
i,j\rangle,\sigma} (a_{i,\sigma}^\dag b_{j,\sigma}+\text{H.c.})
\label{hg}
\end{equation}
where $a_{i,\sigma}^\dag (a_{i,\sigma})$ creates (annihilates) an
electron with the spin $\sigma$ on the position $\mathbf{R}_i$ of
the sublattice $A$, $b_{i,\sigma}^\dag (b_{i,\sigma})$
creates(annihilates) an electron with the spin $\sigma$ on the
position $\mathbf{R}_i$ of the sublattice $B$,
$t_{\text{g}}$ is the nearest neighbor$(\langle
i,j\rangle)$ hopping energy in graphene layer.
Measuring the energy with respect to the Dirac point of the unperturbed graphene
leads to $\epsilon(t)=V_{g}$.

In the momentum space Eq.(\ref{hg})
can be rewritten as
\begin{equation}
\begin{array}{cll}
H_G=\sum\limits_{\mathbf{q},\sigma}
[\epsilon(t)(a_{\mathbf{q},\sigma}^\dag
a_{\mathbf{q},\sigma}+b_{\mathbf{q},\sigma}^\dag
b_{\mathbf{q},\sigma})\\
+\phi(\mathbf{q})a_{\mathbf{q}\sigma}^\dag
b_{\mathbf{q}\sigma}+\phi(\mathbf{q})^* b_{\mathbf{q}\sigma}^\dag
a_{\mathbf{q}\sigma} ]
\end{array}\label{hg12}
\end{equation}
where $\phi(\mathbf{q})=-t_{\text{g}}\sum\limits_{i=1}^3
e^{i\mathbf{q}\cdot\mathbf{\delta_i}}$ with
$\delta_1=\frac{a}{2}(1,\sqrt{3},0),
\delta_2=\frac{a}{2}(1,-\sqrt{3},0),\delta_3=a(1,0,0)$(here $a$ is
the lattice spacing). Upon diagonalizing the  Hamiltonian (\ref{hg12}) one finds
 $E_{\pm}(\mathbf{q})=\pm
t_{\text{g}}|\phi(\mathbf{q})|$, which can be linearized
 around the $\mathbf{K}$
points of the Brillouin zone with the dispersion given by
\begin{equation}
E_{\pm}(\mathbf{q})=\pm v_F|\mathbf{q}|,
\end{equation}
where $v_F=3t_{\text{g}}a/2$ is the Fermi velocity of electron.
From Eq.(\ref{hg12}), one infers that the ac field couples to  the two Dirac
cones equally as a time-dependent gate voltage. Hence,  it does not induce
any transitions between the valleys. Therefore, one only need to
evaluate the contributions of the single valley\cite{peres}, and
then multiply the final results by a factor $2$.
The coupling between the electrodes and the graphene is modeled by
\begin{equation}
H_T=\frac{1}{\sqrt{N}}\sum\limits_{\mathbf{kq}\alpha\sigma}
[T_{\mathbf{k}\alpha\mathbf{q}}c_{\mathbf{k}\alpha\sigma}^\dag
a_{\mathbf{q}\sigma}+ \text{H.c.}],
\end{equation}
where $T_{\mathbf{k}\alpha\mathbf{q}}$ is the coupling matrix
between the $\alpha$ electrode and the graphene; N is the number
of sites on the sublattice $A$.

 Using the nonequilibrium Green's function
method, the electrical current can be expressed as
\begin{equation}
\begin{array}{cll}
I_\alpha(t) &=&-\frac{ie}{\hbar} \int_{-\infty}^t
dt_1\int\frac{d\varepsilon}{2\pi}Tr\{
[(\mathcal{G}_{a}^{r}(t,t_1)-\mathcal{G}_{a}^{a}(t,t_1))f_L(\varepsilon)\\
&&+\mathcal{G}_{a}^{<}(t,t_1)]\Gamma_{L}(\varepsilon)\}
e^{-i\varepsilon(t_1-t)}\\
\end{array}\label{jl3}
\end{equation}
where $Tr$ is the trace in the spin space, $f_\alpha(\varepsilon)$
is Fermi distribution function,
$\mathcal{G}_{a}^{r}(t,t)=\sum\limits_{\mathbf{qq}'}G_{\mathbf{q}a,\mathbf{q}'a}^{r}(t,t')$
and
$\mathcal{G}_{a}^{<}(t,t')=\sum\limits_{\mathbf{qq}'}G_{\mathbf{q}a,\mathbf{q}'a}^{<}(t,t')$
are  $2\times 2$ matrices representing  the retarded green's
function and the lesser Green's function, respectively. In the
calculation of Eq.(\ref{jl3}), we assume that the dominant
contributions to tunneling stem from the electrons near Fermi
level, and hence assume the linewidth function to be  independent
of $\mathbf{q}$. Thus, we have
\begin{equation}
\Gamma_{\alpha }=\left(
\begin{array}{cc}
\Gamma_{\alpha }^\uparrow & 0 \\
0 & \Gamma_{\alpha }^\downarrow%
\end{array}%
\right)
\end{equation}
with $\Gamma_\alpha^\sigma =2\pi\sum\limits_\mathbf{k}
T_{\mathbf{k}\alpha\mathbf{q}}^*T_{\mathbf{k}\alpha\mathbf{q'}}
\delta(\varepsilon-\varepsilon_{\mathbf{k}\alpha\sigma})$.

To calculate $\mathcal{G}_a^{r,a}(t,t')$ in
Eq.(\ref{jl3}) we carry out the gauge transformation
\begin{equation}
\mathcal{G}_{a}^{r,a}(t,t')=\widetilde{\mathcal{G}}_a^{r,a}(t,t')e^{-i\int_{t'}^t
dt_1 \epsilon(t_1)}\label{gqaqa}
\end{equation}
and substitute  in the equation of motion. One  obtains then
\begin{equation}
\begin{array}{cll}
\widetilde{\mathcal{G}}_a^{r,a}(t,t')
&=&\int \frac{d\varepsilon}{2\pi}
\overline{g}_{a}^{r,a}(\varepsilon)[1-\overline{g}_{a}^{r,a}(\varepsilon)\Sigma^{r,a}(\varepsilon)
]^{-1}e^{-i\varepsilon (t-t')}
\end{array}\label{garatt}
\end{equation}
where
$\overline{g}_{a}^{r,a}(\varepsilon)=\frac{1}{N}\sum\limits_\mathbf{q}
g_{\mathbf{q}a,\mathbf{q}a}^{r,a}(\varepsilon)$ and $
\Sigma^{r,a}= \mp\frac{i}{2}[\Gamma_{L}(\varepsilon) +R
\Gamma_{R}(\varepsilon)R^\dag] $ with
$g_{qa,qa}^{r,a}(\varepsilon)=\frac{\varepsilon}{(\varepsilon\pm
i\eta)^2-|\phi(q)|^2}$, and
$$R=\left(
\begin{array}{cc}
\cos \frac{\theta }{2} & -\sin \frac{\theta }{2} \\
\sin \frac{\theta }{2} & \cos \frac{\theta }{2}%
\end{array}%
\right) .
$$
 Introducing a cutoff $k_c$ leads to
\begin{equation}
\overline{g}_{a}^{r}(\varepsilon)=-F_0(\varepsilon)-i\pi\rho_0(\varepsilon),\label{gari}
\end{equation}
\begin{equation}
F_0(\varepsilon)=\frac{\varepsilon}{D^2}\ln\frac{|\varepsilon^2-D^2|}{\varepsilon^2},\
\
\rho_0(\varepsilon)=\frac{|\varepsilon|}{D^2}\theta(D-|\varepsilon|)
\end{equation}
 with $D=v_Fk_c$ denoting a high-energy cutoff of the graphene bandwidth.
$k_c$ is chosen as to guarantee the conservation of the total
number of states in the Brillouin zone after the linearization of
the spectrum around the $K$ point, this is achieved  following the
Debye's prescription.

The lesser Green function $\mathcal{G}_{a}^{<}(t,t')$ can be
derived by applying the analytic continuation rules (cf. Ref.\onlinecite{lang})
 to the equation of motion of the time-ordered Green's
function on a complex contour (Keldysh, Kadanoff-Baym, or another choice of contour),
\begin{equation}
\mathcal{G}_{a}^{<}(t,t')=\int\frac{d\varepsilon_1}{2\pi}\int\frac{d\varepsilon_2}{2\pi}
\widetilde{\mathcal{G}}_a^<(\varepsilon_1,\varepsilon_2)
e^{-i\varepsilon_1t+i\varepsilon_2t'}e^{-\frac{i}{\hbar}\int_{t'}^t\epsilon(t_1)dt_1}\label{selfless}
\end{equation}
where $\widetilde{\mathcal{G}}_{a}^{<}(\varepsilon_1,\varepsilon_2
)=\widetilde{\mathcal{G}}_{a}^{r}(\varepsilon_1 )\Sigma
^{<}(\varepsilon_1,\varepsilon_2
)\widetilde{\mathcal{G}}_{a}^{a}(\varepsilon_2 )$ with
$\widetilde{\mathcal{G}}_{a}^{r,a}(\varepsilon )$ denoting the
Fourier transformation of Eq.(\ref{garatt}), and
\begin{equation}
\begin{array}{cll}
\Sigma^<(\varepsilon,\varepsilon')&=&i2\pi\sum\limits_{mn}J_m(\frac{V_{ac}}{\omega})J_n(\frac{V_{ac}}{\omega})
\\

&&\times \Gamma_{L
}f_L(\varepsilon+m\omega)\delta[\varepsilon-\varepsilon'+(m-n)\omega]

\\

&&+i2\pi\sum\limits_{mn}J_m(\frac{V_{ac}}{\omega})J_n(\frac{V_{ac}}{\omega})
\\

&&\times R\Gamma_{R }R^\dag
f_R(\varepsilon+m\omega)\delta[\varepsilon-\varepsilon'+(m-n)\omega]
 \\
\end{array}
\end{equation}
where $J_m$ is the $m$th order Bessel function of the first kind.
The identity $e^{i\alpha\sin(\omega
t)}=\sum_{m=-\infty}^{\infty}J_m(\alpha)e^{im\omega t}$ is used in
the calculation of Eq.(\ref{selfless}).

Substituting Eqs.(\ref{gqaqa}) and (\ref{selfless}) in
Eq.(\ref{jl3}),
 we finally obtain the time-averaged current
 \begin{equation}
\begin{array}{cll}
 I&=&\frac{e}{\hbar} \sum\limits_{m}
J_{m}^2(\frac{V_{ac}}{\omega})
\int\frac{d\varepsilon}{2\pi}\text{Tr} \{
\mathcal{G}_{a}^{r}(\varepsilon)R\Gamma_{R }R^\dag
\mathcal{G}_{a}^{a}(\varepsilon)\Gamma_{L}\}\\
&&\times[f_R(\varepsilon+m\omega)-f_L(\varepsilon+m\omega)].
\end{array}\label{jL5}
\end{equation}
which is an exact response of the system without imposing any
restriction on the amplitude of the external electric field
$V_{ac}$. For a weak ac field ($V_{ac}<<\omega$),
$J_m(\frac{V_{ac}}{\omega})\approx\frac{1}{\Gamma(m+1)}(\frac{V_{ac}}{2\omega})^m$.
Thus, in this case, the contributions to the tunneling from  the high sidebands  is
suppressed.  While for the large amplitude of the ac field
($V_{ac}>>\omega$),
$J_m(\frac{V_{ac}}{\omega})\approx\sqrt{\frac{2\omega}{\pi
V_{ac}}}\cos(\frac{V_{ac}}{\omega}-\frac{m\pi}{2}-\frac{\pi}{4})$.
Thus,  more sidebands  contribute then to the
transport (we note however, that we are assuming that the external ac field is
harmonic. For this reason we cannot describe with the present method the case of very short pluses
in which case the harmonics associated with the pulse width become also relevant).
 In Eq.(\ref{jL5}), we further set the symmetrical
voltage
 division:$\mu_{L,R}=E_F\pm\frac{1}{2}eV$, and put $E_F=0$ in the numerical calculations.
The TMR can be obtained according to the
  conventional definition
\begin{equation}
\text{TMR}=\frac{I(0)-I(\pi)}{I(0)},\label{tmr}
\end{equation}
where $I(0,\pi)$ is the time-averaged current flowing through the
system in the parallel (antiparallel) configuration.
\begin{figure}[h]
\includegraphics[width=0.8\columnwidth ]{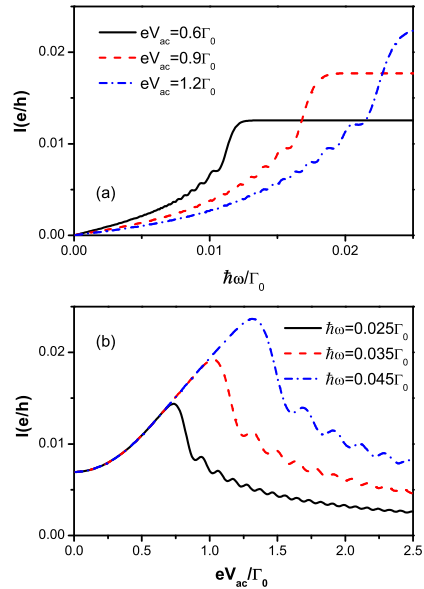}
\caption{(color online) The averaged current $I$ as a function of
the ac frequency for the different ac strength (a), and the ac
strength for the different frequency (b) for parallel
configuration of the electrodes magnetizations. The other
parameters are taken as $k_BT=0.001\Gamma_0, P=0.4,D=8\Gamma_0$,
$eV=\Gamma_0$, where $\Gamma_0$ stands for the coupling between
the scattering region and
 the electrodes.
 }\label{fig1}
\end{figure}

\section{Numerical analysis}
Before performing  and discussing numerical calculations  we shall  clarify first the
assumptions specific  to  the  present theory.
Adopting the wide bandwidth approximation for the
graphene spin-valve system we neglect the energy dependence of the
linewidth functions $\Gamma_\alpha^\sigma(\varepsilon)$.
Additionally, we assume that the two electrodes are made of the
same material, thus the degree of the spin polarizations of the
left and the right electrodes defined by $P_L$ and $P_R$ can be
written as
$\Gamma_L^{\uparrow\downarrow}=\Gamma_R^{\uparrow\downarrow}=\Gamma_0(1\pm
P)$ where $\Gamma_0$ describes the coupling between the graphene
and the electrodes in absence of an internal magnetization, and is
taken as the energy scale in the following numerical calculations.

The frequency dependence of the averaged electrical current for
the different 
ac field strength is shown in Fig.\ref{fig1}(a). Some oscillations
of the electrical current with the frequency can be seen. These
oscillation peaks are asymmetric and their magnitudes depend on
the weight of the different side bands (given by the Bessel
functions) and is non-universal. When the frequency increases to
the value $\hbar\omega\approx 0.02eV_{ac}$, the frequency
dependence of the current through the graphene device acts as a
pure resistance which does not vary with the frequency.
Additionally, one can easily observe that
at the low frequency region, the electrical current increases when
the ac strength $V_{ac}$ grows. This result is equivalent to the
one obtained in ferromagnet-insulator-ferromagnet(FM-I-FM) system
\cite{zhuprb2003}. The reason for this coincidence is easy to
understand: the DOS in graphene vanishes at the Dirac point. The
ac field at the low frequency region stimulates the absorption and
emission of photons close to the Dirac point, and thus can not
break the insulator-type properties of graphene. In this case,
graphene sheet can still be viewed as a tunneling barrier similar
to FM-I-FM system. In the high frequency region, the
photon-induced channels away from the Dirac point contribute
predominantly to the transport, thus leading to the increase of
the electrical current with the ac strength. The $V_{ac}$
dependence of the electrical current for  different frequencies is
shown in Fig. \ref{fig1}(b). The current first increases with
$V_{ac}$, and then decreases involving small oscillations when
$eV_{ac}>30\hbar\omega$. The nonmonotonic dependence of the
electrical current with $V_{ac}$ is different from FM-I-FM system.
The reason is that 
the high sideband tunneling is dominant in the
ferromagnet-graphene-ferromagnet(FM-G-FM) system, while the
population probability modulation of the sidebands
$J_m^2(V_{ac}/\omega)$ tends to suppress its contribution.
Hence, their combinations result in the prediction that there will
be a maximum in the current for some ac voltage.

\begin{figure}[h]
\includegraphics[width=0.8\columnwidth ]{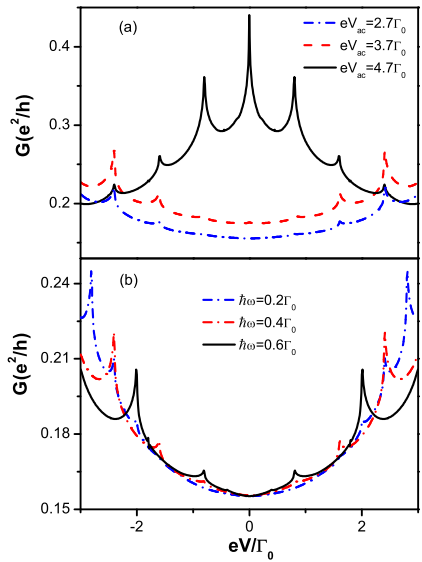}
\caption{ (color online) The dc bias dependence of the
differential conductance $G$ for different ac strength $V_{ac}$ at
$\hbar\omega=0.4\Gamma_0$ (a) and for different frequency $\omega$
at $eV_{ac}=2.7\Gamma_0$ (b) for parallel configuration of the
electrodes magnetizations. The other parameters are taken as those
of Fig.\ref{fig2}.
}\label{fig2}
\end{figure}
 Fig.\ref{fig2}
shows the dc bias dependence of the differential conductance $G=d
I/dV$ for  the different ac strength and frequency $\omega$ in the
parallel electrodes magnetizations.
The differential conductance as a function of the dc voltage
exhibits  successive resonant peaks that correspond to a resonant
tunneling through the photon-induced sidebands. It is interesting
to observe that
 there exists a strong competition between the ac field effect and the Dirac point in graphene.
 For a small ac field amplitude, the zero DOS at the Dirac point of graphene
 suppresses strongly the ac field effect, thus diminishing   the central peak(weighted by $J_0^2$ )
 in the differential conductance versus the bias, as shown in Fig.\ref{fig2}(a). When the ac field strength becomes
 sufficiently large, the insulator-like properties of the Dirac point of graphene are destroyed,
 which resembles the applied magnetic field
case\cite{ding2009prb}. If $eV=0$, the ac field still pumps
electrons through the structure behaving like an effective finite
DOS. This
 leads to the appearance of an implicit central peak in the differential conductance at  zero bias.
 This behavior is a marked difference to the conventional tunneling junction\cite{platpr2004,zhuprb2004,vanarxiv},
 where the central peak is preserved  in the entire range of ac field strength. This characteristic features suggest that
 for the graphene tunneling junction,
 it is possible to externally manipulate the central peak induced by photon in the conductance
  by changing the ac field amplitude.
 Additionally, one can find that the sizes of the side-band peaks  rise  monotonously with the dc bias voltage
 for the small ac field amplitude, but  decrease however for a
 sufficiently large ac field amplitude.
 The reason of this behaviour can be   traced
  back to  a combined effect of the ac field and the nature of
 graphene. For the weak ac field, the linear DOS of the
graphene dominates the transport, and thus modulates
the magnitude of each resonant peak. However, when the ac field
becomes sufficiently strong, the population probability of the
sidebands is suppressed leading to the decrease of the resonant
peaks versus the bias.
Fig.\ref{fig2}(b) shows that with increasing the frequency
$\omega$ of the ac field, the interval between the resonant peaks
increases reflecting an increase of the distance between the
photonic sidebands. The additional phenomenon is that with
increasing the frequency, the each side-band conductance peak
shows a slight rise because of the enhance of the side-band
contribution to the tunneling by its population probability

%
\begin{figure}[h]
\includegraphics[width=0.8\columnwidth ]{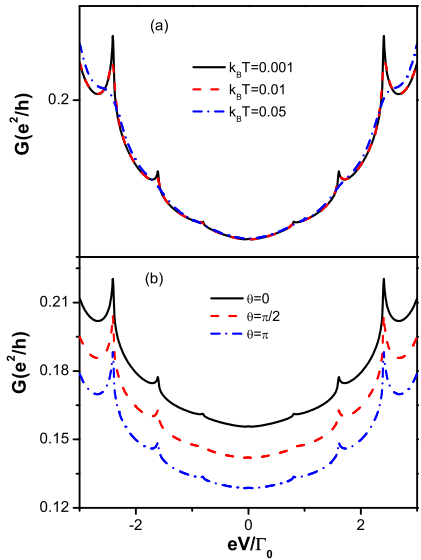}
\caption{(color online)The bias dependence of the differential
conductance $G$ for different temperature $T$  at $\theta=0$,
$eV_{ac}=2.7\Gamma_0$,$\hbar\omega=0.4\Gamma_0$ (a) and for
different angle $\theta$ at $k_BT=0.001\Gamma_0$,
$eV_{ac}=2.7\Gamma_0$, $\hbar\omega=0.4\Gamma_0$ (b). The other
parameters are taken the same as those of Fig.\ref{fig2}.
}\label{fig3}
\end{figure}
The dc bias dependence of the differential conductance for
different temperatures $T$ and angles $\theta$ are shown in
Fig.\ref{fig3}. With increasing the temperature, the resonant
peaks in the differential conductance decrease and almost vanish
at larger $T$. This temperature dependence of the conductance
peaks is similar to that of a noninteracting, single-particle
resonance in the multi-channel model.
The mechanism  is that at high
temperatures not all electron states of the low-lying subbands are
fully occupied  due to the occupation  of the next
 subbands \cite{vanprb1991}. However, near $V=0$, the conductance
is almost independent of the temperature. This characteristic
feature is different from that of the graphene system in the
absence of the ac field. The zero bias conductance in the latter
is sensitive to the temperature \cite{ding2009prb}.  This
insensitivity here can be understood due to a lifting  of the
insulator-type properties of graphene at the Dirac point in the
presence of the ac field. Fig. \ref{fig3}(b) shows
that 
a monotonous suppression of the differential conductance with
increasing the angle $\theta$ takes place in the whole dc voltage
range. This stems from the fact that when $\theta$ changes from 0
to $\pi$ the number of spin-up and spin-down electrons is
rearranged. Therefore the couplings for spin-up and spin-down
electrons become different and the conductance decreases.

\begin{figure}[h]
\includegraphics[width=0.8\columnwidth ]{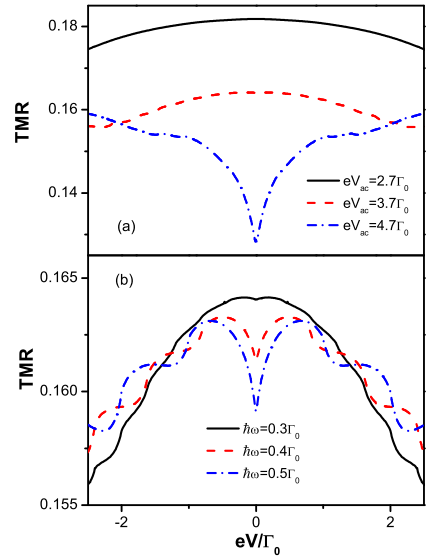}
\caption{(color online)  The dc bias dependence of the TMR for
different ac strength  $V_{ac}$ at $\hbar\omega=0.2\Gamma_0$ (a)
and for different  frequency $\omega$ at $eV_{ac}=3.2\Gamma_0$
(b). The other parameters are taken the same as those of
Fig.\ref{fig2}.}\label{fig4}
\end{figure}

The dc bias dependence of the TMR, defined in Eq.(\ref{tmr}), for
 different ac strengths $V_{ac}$ and  different frequencies
$\omega$ is shown in Fig. \ref{fig4}. At nonzero bias voltage, the
TMR has a step-like structure as a function of the bias voltage
which is caused by  photon-assisted effects.
When increasing the ac field strength, the TMR decreases since the
subbands induced by the ac field enhance the electron transport
through graphene [cf. Fig. \ref{fig2}(a)], however they are spin
independent, thus giving the same contributions to the
spin-dependent transport and leading to the decrease of the TMR
with the ac strength $V_{ac}$. However for the sufficiently large
ac field strength, the TMR at the high bias voltage has a slight
rise which is related to the suppress of the population
probability.
Remarkably, from Fig.\ref{fig4}(a), one can see that the TMR
exhibits a salient transition from a broad peak to a sharp dip at
the zero bias voltage. This is due to the strong competition
between the ac field effect and the Dirac point of graphene.
An effective DOS developing by the large ac field strength
 pushes the central region from the insulator-like to a more
metal-like behaviour. This results in a strong decrease of TMR at
zero bias.
 Additionally, one can find that with increasing the
frequency, the steps in the TMR become broad due to the increase
of the distance between the sidebands, as shown in
Fig.\ref{fig4}(b). In particular, with increasing the frequency,
the amplitude of the dip in the TMR at zero bias increases due to
the lift of the zero bias conductance by  the population
probability of the main sideband.

\section{Summary}
In conclusion, we studied theoretically the spin-dependent
transport through the FM-G-FM system in the presence of an
external ac field by means of the time dependent nonequilibrium
Green's function approach. We obtained  analytic formulas for the
electrical current, and found that there exists a strong
interplay between the ac field effect and the Dirac point in
graphene. For a small ac field amplitude, the zero DOS at the Dirac
point of graphene suppresses strongly the ac field effect, and
diminishes  some of the photon-induced resonant peaks in the differential conductance
 when varied as a function of the
the bias. For a   sufficiently large ac field strength, the
insulator-like properties of the Dirac point of graphene are
lifted which leads to the appearance of prominent resonant  peaks in the
differential conductance, particularly  at a zero bias. In this situation, the
TMR exhibits a transition from a peak to a sharp dip at the zero
bias voltage due to this subtle competition mechanism.  This dip
magnitude can be manipulated even by changing the ac field
frequency. Therefore, it is suggested that for a graphene tunnel
junction, it is possible to externally manipulate the central peak
induced by photons in the conductance and the zero bias TMR by
changing the ac field amplitude or frequency.

\begin{acknowledgments}
The work of K.H.D. is supported by  DAAD (Germany) and by the National Natural Science
Foundation of China (Grant Nos. 10904007), the Natural Science
Foundation of Hunan Province, China (Grant No. 08JJ4002 ), and the
construct program of the key discipline in Changsha University of
Science and Technology, China. J.B. and Z.G.Z. are supported by DFG, Germany.
\end{acknowledgments}


\begin{thebibliography}{99}
\bibitem{footn} Other type of couplings through $\Gamma_{2}$ and $\Gamma_{6}$ representations corresponds to
  different physical situations and are the subject of ongoing work.
\end{thebibliography}
\end{document}